\documentclass[aps,prb,twocolumn,showpacs,superscriptaddress]{revtex4}
\usepackage{graphicx}
\usepackage{dcolumn}
\usepackage{bm}
\usepackage{color}
\usepackage{amsmath}
\usepackage{amssymb}

\begin{document}
\title{Surface plasmon Fourier optics}
\author{A.~Archambault}
\affiliation{Laboratoire Charles Fabry de l'Institut d'Optique,
CNRS and Universit\'e Paris-Sud, Campus Polytechnique, RD 128,
91127 Palaiseau cedex, France}
\author{T.~V.~Teperik}
\affiliation{Laboratoire Charles Fabry de l'Institut d'Optique,
CNRS and Universit\'e Paris-Sud, Campus Polytechnique, RD 128,
91127 Palaiseau cedex, France} \affiliation{On leave from:
Institute of Radio Engineering and Electronics (Saratov Division),
\\ Russian Academy of Sciences, Zelyonaya 38, 410019 Saratov,
Russia}
\author{F.~Marquier}
\affiliation{Laboratoire Charles Fabry de l'Institut d'Optique,
CNRS and Universit\'e Paris-Sud, Campus Polytechnique, RD 128,
91127 Palaiseau cedex, France}
\author{J.J.~Greffet}
\email[Electronic address:
]{jean-jacques.greffet@institutoptique.fr}
\affiliation{Laboratoire Charles Fabry de l'Institut d'Optique,
CNRS and Universit\'e Paris-Sud, Campus Polytechnique, RD 128,
91127 Palaiseau cedex, France}

\begin{abstract}
Surface plasmons are usually described as surface waves with
either a complex wavevector or a complex frequency. When
discussing their merits in terms of field confinment or
enhancement of the local density of states, controversies
regularly arise as the results depend on the choice of a complex
wavevector or a complex frequency. In particular, the shape of the
dispersion curves depends on this choice. When discussing
diffraction of surface plasmon a scalar approximation is often
used. In this work, we derive two equivalent vectorial
representations of a surface plasmon field using an expansion over
surface waves with either a complex wavevector or a complex
frequency. These representations can be used to account for
propagation and diffraction of surface waves. They can also be
used to discuss the issue of field confinment and local density of
states as they have a non-ambiguous relation with the two
dispersion relations.
\end{abstract}
\maketitle

\section{Introduction}

Surface plasmons have been known since the pioneering work of
Ritchie in the 1950s \cite{Ritchie}. Considerable advances made in
nanotechnology in recent years and the desire to control and
manipulate light at nanoscale have renewed the interest in surface
plasmons \cite{Barnes}. Numerical  simulations and experiments
have demonstrated unique properties of different plasmonic
nanostructure such as extraordinary transmission
\cite{Ebbesen1,LalanneN}, guiding\cite{Weeber, Bozhevolnyi, Krenn,
Atwater},  fluorescence enhancement \cite{Barnes2,Novotny,
Sandoghdar, Rigneault, Khurgin}, field
enhancement\cite{Nordlander,Halas, Schatz},
focussing\cite{ZhangNL}, superresolution \cite{Zhang,
Zayats,Taubner}, omnidirectional absorption
\cite{Teperik,Marquier}, coherent thermal
emission\cite{GreffetN,Marquier,Arnold}.

In this paper, we shall focus on surface plasmons propagating
along flat surfaces. Propagation of surface plasmons on a flat
surface perpendicular to the z axis is often discussed using a
mode $E(z)\exp[i(K_xx+K_yy -\omega t)]$ characterized by a
frequency $\omega$ and a wave vector $\mathbf{K} = K_\text{x}
\mathbf{\hat{x}} + K_\text{y} \mathbf{\hat{y}}$ parallel to the
interface. However, the surface plasmon fields diffracted by
edges, guided by ridges, focussed by lenses cannot be described by
a simple mode. It is well-known that a finite size beam
propagating in a vacuum has to be described in terms of a linear
superposition of plane waves. Differents ansatz, often neglecting
polarization, have been used in the litterature to address this
question\cite{Weeber, Fainman,Brongersma}. One of the goals of
this paper is to derive a rigorous representation for the surface
plasmon field. Such a superposition is the equivivalent of the
angular plane wave spectrum for surface plasmons It can be used to
develope a framework for surface plasmon fourier optics.

\begin{figure}
\includegraphics{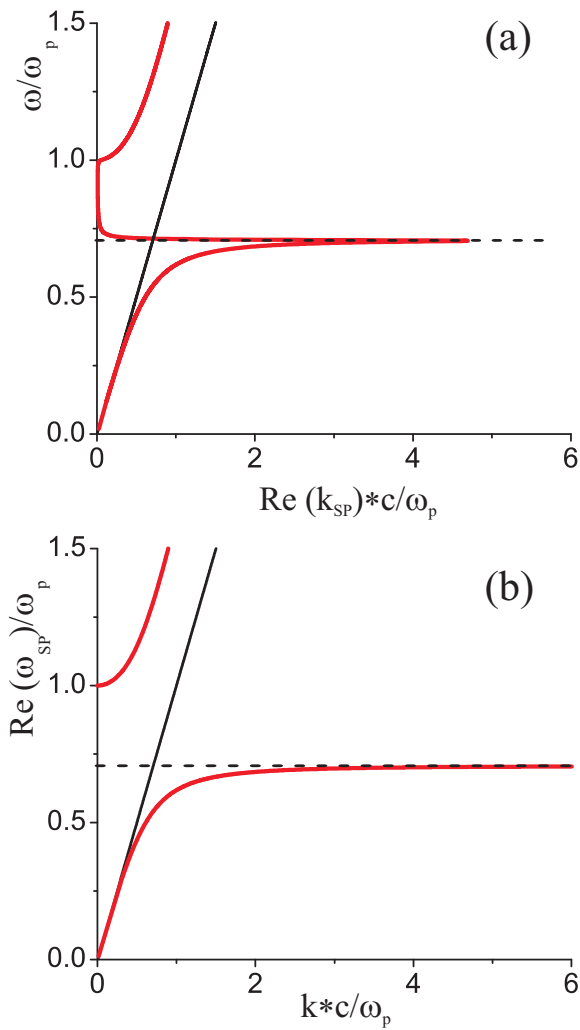}
\caption{\label{dispcurve} (color online) Dispersion of surface
plasmons propagating along the metal/air interface. (a) Real
$\omega$ is chosen to obtain a complex $K_{sp}$. (b) Real $K$ is
chosen to obtain a complex $\omega_{sp}$. The flat asymptote
(dashed line) situated at $\omega_p/\sqrt{2}$ represents the
nonretarded surface plasmon solution. The slanting solid line
represents the light cone inside which a wave is propagating
(radiative) and outside which is evanescent (surface).}
\end{figure}

In doing so, a difficulty arises. When losses are taken into
account, a mode with real $\mathbf{K}$  and real $\omega$ is no
longer a valid solution. Although we can still use a Fourier
representation with  real $\mathbf{K}$ and real $\omega$, it is
not convenient to deal with waves that are not a solution.
Elementary solutions using either a complex $\mathbf{K}$ or a
complex $\omega$ can be found. However, we cannot assume that they
form a basis. The first issue is thus to derive a general
representation for the surface plasmon field as a superposition of
modes. The second issue is related to the dispersion relation. A
dispersion relation can be found when using either a complex
$\mathbf{K}$ and a real $\omega$ or vice versa. These two choices
leads to different shapes as seen in Fig.\ref{dispcurve}. One
dispersion relation has an asymptote for very large values of $K$
while the other has limited values of $K$ and presents a
backbending.

This issue was first noted by Alexander et al.\cite{Alexander} and
later discussed by Arakawa et al. \cite{Arakawa}. Arakawa remarked
that when plotting the position of the dips in a reflectivity
experiment where the angle of incidence is varied at fixed
frequency, one finds the dispersion relation with backbending.
Instead, when plotting the points obtained from a spectrum at
fixed angle, one finds the dispersion relation without
backbending. This approach seems simple and easily applicable. It
is sufficient to explain the attenuated total reflection (ATR)
experiments. Nevertheless, it is not a general prescription that
can be used to discuss all possible issues. Let us illustrate this
point by addressing two questions regarding the most important
properties of surface plasmons~: confinment of the fields and
large density of states. For a theoretical discussion of these
applications, different dispersion relations lead to different
predictions. Confinment of the field is the key property regarding
applications such as optical lithography, enhanced non-linear
effects or super-resolution issues. The dispersion relation with a
backbending predicts a cut-off spatial frequency and therefore a
resolution limit whereas the dispersion relation without
backbending does not predict any resolution limit. Enhancement of
the local density of states (or Purcell effect) is fundamental for
fluorescence enhancement and more light emission assisted by
surface plasmons. The dispersion relation with a backbending again
predicts a cut-off spatial frequency and therefore an upper limit
to the LDOS. No limit is predicted by the other dispersion
relation. A general discussion on the applicability of the
different dispersion relations is thus needed.

In this paper we start by deriving two general representations of
the surface plasmon field in terms of linear superpositions of
modes having the structure $\exp[i(K_xx+K_yy+\gamma z-\omega t)]$
with a well-defined polarization. Each representation is
associated with either a complex frequency or a complex wavevector
and therefore to a particular dispersion relation. We then show
that the most convenient choice depends on the physical problem to
be discussed. We introduce a prescription that allows to choose
complex or real frequency and the associated dispersion relation.
We then apply our analysis to discuss the resolution limit issue
and the LDOS issue. The paper is organized as follows. For the
sake of completeness, we briefly summarize the derivation of the
dispersion relation in the next section. The following section
introduces the general representations of the surface plasmon
field. We then discuss the physical content of these
representations.

\section{Surface plasmon dispersion relation}

Let us consider a flat metal surface $z=0$ bounded by dielectric
media with dielectric constant $\varepsilon_1$. For convenience,
we describe the dielectric response of the metal to an electric
field using the local Drude model
\begin{equation}
\epsilon_2(\omega)=1-\frac{\omega_\text{p}
^2}{\omega(\omega+i\nu_e)}, \label{epsilon}
\end{equation}
where $\omega_p$ is the bulk plasmon frequency and $\nu_e$ is a
phenomenological bulk electron relaxation rate. We derive the
dispersion relation of surface plasmons propagating along the
metal/dielectric interface.

We search a solution of Maxwell equations for an interface between
two linear isotropic and local media characterized by dielectric
constants $\epsilon_m$ where $m=1,2$ denotes medium $1$ ($z<0$) or
$2$ ($z>0$). A surface wave solution has a structure
$\exp[i(\mathbf{K}\cdot\mathbf{r}+\gamma _m \lvert z \rvert
-\omega t)]$ with
\begin{equation}
\mathbf{K}^2+\gamma _m ^2=\epsilon_m\omega^2/c^2
\end{equation}
where $\gamma _m$ is chosen so that $\mathfrak{Im} \gamma _m > 0$.
Boundary conditions impose the continuity of the tangential components of the electric field and $\epsilon E_z$. It follows that a p-polarized field can exist provided that $\epsilon_1\gamma_2=-\epsilon_2\gamma_1$. One finds that a solution is given by
\begin{equation}
K^2=\left(\frac{\omega}{c}\right)^2\frac{\epsilon_1\epsilon_2}{\epsilon_1+\epsilon_2}.
\label{reldisp}
\end{equation}
When dealing with an interface separating a dielectric from a non
lossy metal,
$\frac{\displaystyle\epsilon_1\epsilon_2}{\displaystyle\epsilon_1+\epsilon_2}$
yields a unique solution to the problem. When accounting for
losses in the material,
$\frac{\displaystyle\epsilon_1\epsilon_2}{\displaystyle\epsilon_1+\epsilon_2}$
is a complex number so that the dispersion relation cannot be
solved using real $K$ and real $\omega$. It is necessary to
consider a complex frequency or a complex wavector to find roots
of the equation. Let us first choose $\omega$ real. We denote
$K_{sp}$ the complex root of the equation \eqref{reldisp}.
Fig.\ref{dispcurve}(a) shows the dispersion curve obtained from
the surface plasmon dispersion relation Eq.~(\ref{reldisp}) when
plotting $\omega$ versus $\mathfrak{Re}K _{sp}$. This curve
exhibits a back-bending in the vicinity of the frequency of
non-retarded surface plasmon $\omega_p/\sqrt{2}$.  The second
possible choice is to keep a real wavector $\mathbf{K}$. We denote
$\omega_{sp}$ the complex root of the equation
Eq.~(\ref{reldisp}). Fig.\ref{dispcurve}(b) shows the dispersion
curve obtained when plotting $\mathfrak{Re} \omega_{sp}$ versus
$K$. It is seen in Fig.\ref{dispcurve}(b) that this curve exhibits
an asymptote  for large wavevectors.

Let us make two remarks regarding the dispersion relation. We
first note that Eq.(\ref{reldisp}) is also a solution of
$\epsilon_1\gamma_2=\epsilon_2\gamma_1$ which defines a zero of
the reflection factor, i.e. the Brewster angle. It can be checked
that the upper branch in Fig.\ref{dispcurve}(b) is not a surface
wave but the locus of the Brewster angle in the $(\omega,K)$
plane. Finally, we note that a surface plasmon is a collective
oscillation of charge density. When the frequency $\omega$ is
smaller than $\nu_e$, the collective electron oscillation is
overdamped. In this frequency regime, Eq.(\ref{reldisp}) describes
a surface wave that has no longer the character of a surface
plasmon. Thus, Eq.(\ref{reldisp}) describes the Brewster angle for
$\omega>\omega_p$, a surface plasmon for $\omega_p>\omega>\nu_e$,
a surface wave for $\nu_e>\omega$.

\section{General field representations}

The aim of this section is to derive a general form of the surface
plasmon field. To this aim, we first use the simple interface
Green's tensor that yields the general form of the field for any
given source distribution. We then extract the surface plasmon
contribution which is defined to be the pole contribution to the
Green's tensor. We will show that this procedure leads in a
natural way to different representations that make use of either a
complex wavevector or a complex frequency. We emphasize that both
representations will describe the same electromagnetic surface
plasmon field $\mathbf{E}_{sp}(x,y,z,t)$.

Let us suppose that an arbitrary source is located nearby the
dielectric-metal interface. The electric field generated by the
source $\mathbf{j}(\mathbf{r}, t)$ is given by the relation
\begin{equation}
\mathbf{E}(\mathbf{r},t)= -\mu _0\int\mathrm{d}t'\int\mathrm{d}^3
\mathbf{r}'\,\,\tensor{G} (\mathbf{r},\mathbf{r}',t-t')
\frac{\partial \mathbf{j} (\mathbf{r}',t')}{\partial t'},
\label{field}
\end{equation}
where $\mu _0$ is vacuum permeability. A Fourier representation
can be written in the form
\begin{multline}
\tensor{G} (\mathbf{r},\mathbf{r}',t-t')=\int \frac{\mathrm{d} ^2
\mathbf{K}}{4\pi^2} \int
\frac{d\omega}{2\pi}\tensor{g}(\mathbf{K},z,z',\omega)\\
e^{i[\mathbf{K}(\mathbf{r}-\mathbf{r}')-\omega (t-t')]},
\label{FourG} \end{multline} Here, the integration variables
$\omega$ and $K_x,K_y$ are real. The explicit form of the Green's
tensor $\tensor{g}(\mathbf{K},z,z',\omega)$ in the presence of the
interface is given in Appendix A. It is seen that the Fourier
transform of the Green's tensor has poles given by the denominator
of the Fresnel factors for $p$-polarized field. For a
dielectric/metal interface, they correspond to the surface plasmon
as discussed previously. Thus, the Green's tensor can be split
into two terms : the pole contribution that yields the surface
plasmon and the remaining contribution that yields a regularized
Green's tensor.
\begin{equation}
\tensor{G} = \tensor{G} _{reg} + \tensor{G} _{sp},
 \label{green}
\end{equation}
where the pole contribution to the Green's tensor
$\tensor{G}_{sp}$ can be explicitly derived using the residue theorem.
   $\tensor{G} _{reg}$ is the contribution
of the regularized Green's dyadic. It can be shown that the
Green's tensor can be evaluated using a contour deformation in the
complex plane and that the regularized term is essentially due to
the contribution along the branch cut. This contribution is often
termed cylindrical wave or creeping wave. The relative importance
of these terms is well documented in classical texts for
radiowaves\cite{Banos, Felsen}. The analysis of their respective
contribution was of practical importance in the early days of
telecommunications as radiowaves were guided by the earth. This
issue has been discussed recently in the context of
optics\cite{Lalanne,LalanneN}. In this paper, we shall not pursue
this discussion and focus instead on the surface wave contribution
defined as the pole contribution.
\begin{equation}
\mathbf{E}_{sp}(\mathbf{r},t)=-\mu_0
\int\mathrm{d}t'\int\mathrm{d}^3 \mathbf{r}'\,\,\,\tensor{G}_{sp}
(\mathbf{r},\mathbf{r}',t-t') \frac{\partial \mathbf{j}
(\mathbf{r}',t')}{\partial t'}. \label{fieldsp}
\end{equation}

When solving Eq.(\ref{reldisp}), we can consider that $\omega $ is
real and find a complex $K_{\text{sp}}$ or we can impose a real
value to $K$ and find a complex root $\omega_{sp}$. Thus, when
extracting the poles, it is a matter of choice to consider that
they are poles in the complex frequency plane or in the complex
wavevector plane. We find either a couple of poles  $\omega_{sp}$
and $-\omega^{\ast}_{sp}$ or a complex wavevector pole $K^2 _{sp}$
hence two poles for the component of the wavevector along the $x$
axis $K _{\text{x,sp}}$ and $-K _{\text{x,sp}}$ for a given
component along the $y$ axis $K_y$ as $K^2 _{sp} = K_{\text{x,sp}}
^2 + K_y ^2$. It follows that we can cast the pole contribution to
the Green's tensor in the form :
\begin{equation}
    \tensor{g} _{sp}
    (\mathbf{K}, z, z', \omega) = \frac{\tensor{f}_{\omega_{sp}} ( \mathbf{K}, z, z')}{\omega - \omega_{sp}} +
    \frac{\tensor{f}_{-\omega_{sp}^{\ast}} (\mathbf{K}, z, z')}{\omega + \omega_{sp}^{\ast}},
 \end{equation}
where $\tensor{f}_{\omega_{sp}} ( \mathbf{K}, z, z')$ and
$\tensor{f}_{-\omega_{sp} ^{\ast}} (\mathbf{K}, z, z')$ are the
residues of $\tensor{g}$ at $\omega_{sp}$ and $-\omega_{sp}
^{\ast}$ respectively, or in the form :
\begin{eqnarray*}
&&  \tensor{g}_{sp} (\mathbf{K}, z, z', \omega) = \\
    &&  \qquad\frac{\tensor{f}_{K_{\text{x, sp}}} (K_{\text{y}}, z, z', \omega)}{K_{\text{x}} - K_{\text{x, sp}}}
                    + \frac{\tensor{f}_{-K_{\text{x, sp}}} (K_{\text{y}}, z, z', \omega)}{K_{\text{x}} + K_{\text{x, sp}}}
\end{eqnarray*}
where $\tensor{f}_{K_{\text{x, sp}}} (K_{\text{y}}, z, z',
\omega)$ and $\tensor{f}_{-K_{\text{x, sp}}} (K_{\text{y}}, z, z',
\omega)$ are the residues of $\tensor{g}$ at $K_{\text{x, sp}}$
and $-K_{\text{x, sp}}$ respectively.

These two choices leads to two different forms of the surface
plasmon field given by Eq.(\ref{fieldsp}). We now examine these
forms in detail.

\subsection{Surface plasmon field representation with a real wavevector}
In this section we derive the analytical form of the surface
plasmon field using real wavevectors.  For this purpose we
evaluate the pole contribution to the Green's tensor by
integrating in the complex $\omega$ plane.  The complex pole
$\omega_{sp}$ then yields a contribution for $t-t'>0$ that varies
as $\exp(-i\omega_{sp}(t-t'))$.  After integration, we find :
\begin{multline} \label{G_w}
\tensor{G}_{sp} = H(t-t') \ 2 \mathfrak{Re}  \int
\frac{\mathrm{d}^2 \mathbf{K} }{(2\pi)^2}(-i) \tensor{f} _{\omega
_{sp}} (\mathbf{K}, z, z')\\
 e^{i[ \mathbf{K} \cdot
(\mathbf{r} - \mathbf{r}')- \omega_{sp} (t - t')]}\\
\end{multline}
where $\tensor{f} _{\omega _{sp}} (\mathbf{K}, z, z')$ is the
residue of $\tensor{g}$ at $\omega _{\text{sp}}$. It is given in
the appendix \ref{field_calc}. It follows from Eq.(\ref{fieldsp})
that the field can be cast in the form of a linear superposition
of modes with real wavevector and complex frequency :
\begin{multline}
    \mathbf{E}_{\text{sp}} = 2 \mathfrak{Re}  \int \frac{\mathrm{d}^2 \mathbf{K} }{(2\pi) ^2}\,
     E(\mathbf{K} , t)(\hat{\mathbf{K}} - \tfrac{K}{\gamma _{\text{m}}} \mathbf{n} _{\text{m}} )\\
     e^{i ( \mathbf{K} \cdot \mathbf{r} + \gamma_{\text{m}}
    \lvert z \rvert - \omega_{sp} t ) },
    \label{fieldw}
\end{multline}
where the amplitude $E (\mathbf{K}, t)$ is given in appendix
\ref{field_calc}, $\mathbf{n} _{\text{m}}= -\mathbf{\hat{z}}$ if
$z<0$ and $\mathbf{\hat{z}}$ if $z>0$, and $\mathbf{\hat{K}} =
\mathbf{K} / K$. The surface plasmon field takes a form that looks
as a mode superposition \textit{except that the amplitude $E
(\mathbf{K}, t)$ depends on the time $t$}. Indeed, when describing
a stationary field using modes that have an exponential decay, the
amplitude is necessarily time dependent. In order to obtain a
superposition of modes with fixed amplitudes, it is necessary to
assume that all sources are extinguished after time $t=0$ so that
we observe the field after it has been excited. In that case, the
decay of the mode is well described by the imaginary part of
$\omega_{sp}$. Eq.~(\ref{fieldw}) is thus well suited for fields
excited by pulses. Note that the polarization of each mode is
specified by the complex vector $\hat{\mathbf{K}} -
\tfrac{\displaystyle K}{\displaystyle\gamma _{\text{m}}}
\mathbf{n} _{\text{m}}$, whose component along the $z$ axis
depends on the medium from which the field is evaluated.

\subsection{Surface plasmon field representation with a real frequency}

Let us now turn to the alternative choice. We consider the complex
poles $K_{\text{x, sp}}$ and $-K_{\text{x, sp}}$. The Green
function can be cast in the form~:
\begin{multline} \label{G_Kx}
\tensor{G_{sp}} =i \int\frac{d\omega}{2\pi}\int \frac{\mathrm{d}
K_y }{2\pi}\,\tensor{f} _{K_{\text{x, sp}}} (K_y, z, z', \omega)\\
e^{i K_{\text{x, sp}}(x-x')}e^{i K_y (y-y')} e^{-i \omega (t -
t')}
\end{multline}
if $x-x' > 0$, and~:
 \begin{multline} \label{G_mKx}
\tensor{G_{sp}} =-i\int\frac{d\omega}{2\pi}\int \frac{\mathrm{d}
K_y }{2\pi}\, \tensor{f} _{-K_{\text{x, sp}}} (K_y, z, z',
\omega)\\ e^{-i K_{\text{x, sp}}(x-x')}e^{i K_y (y-y')} e^{-i
\omega (t - t')}
\end{multline}
if $x-x'< 0$. $\tensor{f} _{K_{\text{x, sp}}} (K_y, z, z',
\omega)$ and $\tensor{f} _{K_{-\text{x, sp}}} (K_y, z, z',
\omega)$ are the residues of $\tensor{g}$ at $K_{\text{x, sp}}$
and $-K_{\text{x, sp}}$. They are given in appendix
\ref{field_calc}. When inserting this form in Eq.(\ref{fieldsp}),
we again obtain a form for the field that is a superposition of
modes whose amplitude depends on $x$ :
\begin{multline}
\mathbf{E} \label{fieldKx}
    =
     \int \frac{\mathrm{d}\omega }{2\pi}\, \int \frac{\mathrm{d} K _{\text{y}} }{2\pi}\, \Big[
     E _{>} (K _{\text{y}} , \omega, x) (\mathbf{\hat{K}} ^+ - \tfrac{K _{sp}}{\gamma _\text{m}} \mathbf{n} _\text{m} )\\
     e^{i ( K _{\text{x, sp}}   x + K _{\text{y}} y + \gamma _\text{m} \lvert z \rvert - \omega t ) } \\
     +E _{<}  (K _{\text{y}} , \omega, x)(\mathbf{\hat{K}} ^- - \tfrac{K _{sp}}{\gamma _\text{m}} \mathbf{n} _\text{m} )\\
     e^{i (-K _{\text{x, sp}}   x + K _{\text{y}} y +  \gamma _\text{m} \lvert z \rvert - \omega t ) }
     \Big]
\end{multline}
where $\mathbf{\hat{K}} ^+ = (K_\text{x,sp} \mathbf{\hat{x}} +
K_\text{y} \mathbf{\hat{y}}) / K _{sp}$ and $\mathbf{\hat{K}} ^- =
(-K_\text{x,sp} \mathbf{\hat{x}} + K_\text{y} \mathbf{\hat{y}}) /
K _{sp}$. The amplitudes $E _{>} (K _{\text{y}} , \omega, x)$ and
$E _{<}  (K _{\text{y}} , \omega, x)$ are given in appendix
\ref{field_calc}. Again, it seems natural to have amplitudes of
the modes that depend on $x$ if one describes a homogeneous field
using modes with a decay along $x$. A proper mode representation
should use only fixed amplitudes. This is possible if all the
sources lie in the $x<0$ region and the region of interest is the
$x>0$ region. We then obtain a surface plasmon field that can be
cast in the form :
\begin{multline}
\mathbf{E}
    =
     \int \frac{\mathrm{d}\omega }{2\pi}\, \int \frac{\mathrm{d} K _{\text{y}} }{2\pi}\,
    (\mathbf{\hat{K}} - \tfrac{K _{sp}}{\gamma _\text{m}} \mathbf{n} _\text{m} )
    E _{>} (K _{\text{y}} , \omega)\\
    e^{i ( \mathbf{K}\cdot\mathbf{r} + \gamma _\text{m} \lvert z \rvert - \omega t ) }
    \label{FourOptSP}
\end{multline}
where $\mathbf{K} = K _{\text{x, sp}}  \mathbf{\hat{x}} + K
_{\text{y}} \mathbf{\hat{y}} $ is complex and $\mathbf{\hat{K}} =
\mathbf{K}/{K _{sp}}$. We conclude that stationary monochromatic
fields with a finite size are well described by a representation
that uses complex wavevectors and real frequencies. This equation
is one of the main result of this paper. Indeed, it provides a
framework to develop surface plasmon Fourier optics. Similar
representations have been postulated as ansatz to surface plasmons
interferences \cite{Brongersma}, propagation along a
stripe\cite{Weeber} or focussing\cite{Fainman}. The framework
introduced above provides a rigorous derivation of the form of the
surface plasmon field valid in a region with no sources. Let us
emphasize that this representation is well suited to discuss
propagation for $x>x_0$ of a surface plasmon field known along a
line $x=x_0$. It is seen on Eq.\eqref{FourOptSP} that propagation
over a distance $d$ amounts to multiply each mode by a factor
$\exp(iK_xd)$. In general, this involves modifying both the phase
and the amplitude of the mode. Thus, it allows to discuss any
surface wave diffraction problem. Finally, let us stress that this
representation is valid for  a complex wavevector $\mathbf{K}$ and
a real frequency $\omega$ so that this representation is
necessarily associated with a dispersion relation with
backbending.

To summarize, we have shown that the surface plasmon field can be
represented using modes that have either a complex frequency or a
complex wavevector. However, the amplitudes may still depend on
either time or space. In the case of a field excited by a pulse,
the representation that uses a complex frequency is well suited.
It is associated with the dispersion relation without backbending.
In the case of a stationary monochromatic excitation localized in
space, a representation using modes with complex wavevectors is
well suited. It corresponds to a dispersion relation with
backbending. This simple analysis yields a simple prescription to
choose the proper dispersion relation. Note that in the case of
pulses limited in space, both representations can be used.

\subsection{Surface plasmon field generated by a dipole}

\begin{figure}
\includegraphics{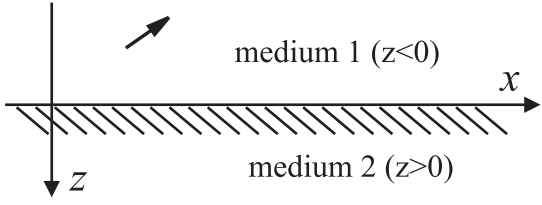}
\caption{ \label{dipfig} A point-like dipole located nearby the
dielectric-metal interface}
\end{figure}

For many applications, it is useful to know the field generated by
a dipole. For instance, when considering the field scattered by a
subwavelength particle, the source can be represented by an
electric dipole. In addition, any source can be decomposed as
linear superposition of dipolar sources. Here, we derive the
surface plasmon field generated by a monochromatic point-like
dipole characterized by its dipole moment $\mathbf{p_0}$ (see Fig.
\ref{dipfig}). Note that there are other contributions to the
field generated by the dipole at a distance typically smaller than
a wavelength. The surface plasmon contribution is typically
dominant for larger distances \cite{Greffet94,Lalanne}.

For a vertical dipole $\mathbf{p_0} e^{-i \omega _0 t} = p_0 e^{-i
\omega _0 t} \mathbf{\hat{z}}$ located at a distance $d$ below the
interface, at $x=y=0$, we obtain, using the cylindrical basis
$\mathbf{\hat{\rho}}, \mathbf{\hat{\theta}}, \mathbf{\hat{z}})$~:
\begin{multline} \label{vertdip}
    \mathbf{E_{\rm m}} =
        2 \mathfrak{Re} \Big\lbrace
            \left[
                H_1^{(1)} (K_{sp} \rho) \mathbf{\hat{\rho}}
                + i \tfrac{\displaystyle K_{sp} }{\displaystyle\gamma _m} H_0^{(1)} (K_{sp} \rho) \mathbf{n} _m
            \right]
            \\
            M(K_{sp} , \omega _0)
            \frac{1}{\epsilon _0}
            \left(-i \tfrac{\displaystyle K_{sp} }{\displaystyle \gamma _1} \right) p_0
e^{i \gamma_1 d} e^{i \gamma _m \lvert z \rvert }
            e^{-i \omega _0 t}
        \Big\rbrace
\end{multline}
where $H_0^{(1)}$ and $H_1^{(1)}$ are Hankel functions of the
first kind of zero-th and first order respectively, $K_{sp} $ is
complex and verifies Eq. \eqref{reldisp} with $\omega = \omega
_0$, $M(K_{sp} , \omega _0)$ is given in appendix \ref{dip_field}
and the other notations are defined above. The details of the
calculation are given in appendix \ref{dip_field}. Using the
asymptotic forms of the Hankel functions, we obtain for the field
of a vertical dipole, for $\rho$ greater than a few $1/\lvert
K_{sp} \rvert$~:
\begin{multline} \label{vertdips}
    \mathbf{E_{\rm m}} =
        2 \mathfrak{Re} \bigg[
            \frac{e^{i K_{sp} \rho}}{\sqrt{K_{sp} \rho}}
            \left(
                \mathbf{\hat{\rho}} - \tfrac{\displaystyle K_{sp} }{\displaystyle\gamma _m} \mathbf{n} _m
            \right)
            M' _v (K_{sp} , \omega _0) \, p_0 \\ e^{i \gamma_1 d} e^{i \gamma _m \lvert z \rvert }
            e^{-i \omega _0 t}
        \bigg]
\end{multline}
where $M' _v (K_{sp} , \omega _0)$  is given in appendix
\ref{dip_field}. Thus the surface plasmon field is analogous to a
damped cylindrical wave $\frac{\displaystyle e^{iK_{sp}
\rho}}{\displaystyle\sqrt{K_{sp} \rho}}$ with a polarization
vector $\mathbf{\hat{r}} - \tfrac{\displaystyle K_{sp}
}{\displaystyle\gamma _m} \mathbf{n} _m$. For a dipole oriented
along the $x$ axis $\mathbf{p}_0 e^{-i\omega_0 t} = p_0
e^{-i\omega_0 t} \mathbf{\hat{x}}$, we obtain~:
\begin{multline} \label{horizdip}
    \mathbf{E_{\rm m}} =
        2 \mathfrak{Re} \bigg\lbrace
            \Big[ \left(
                [H_1^{(1)}(K_{sp} \rho)]' \mathbf{\hat{\rho}}
                 - i \tfrac{\displaystyle K_{sp} }{\displaystyle\gamma _m}
                 H_1^{(1)} (K_{sp} \rho) \mathbf{n} _m
            \right) \\
            \qquad \qquad
            \qquad \qquad
            \cos \theta
            - \frac{H_1^{(1)} (K_{sp} \rho)}{K_{sp} \rho}
            \mathbf{\hat{\theta}} \sin \theta \Big]
            \\
            M(K_{sp} , \omega _0) \
            \frac{1}{\epsilon _0}\,p_0 \,e^{i \gamma_1 d}
            e^{i \gamma _m \lvert z \rvert }
            e^{-i \omega _0 t}
        \bigg\rbrace
\end{multline}
where $\theta = (\mathbf{\hat{x}}, \mathbf{\hat{\rho}})$. Using
the asymptotic forms of the Hankel functions, we obtain~:
\begin{multline} \label{horizdips}
    \mathbf{E_{\rm m}} =
        2 \mathfrak{Re} \bigg[
            \frac{\displaystyle e^{i K_{sp} \rho}}{\displaystyle\sqrt{K_{sp} \rho}}
            \left(
                \mathbf{\hat{\rho}} - \tfrac{\displaystyle K_{sp} }{\displaystyle\gamma _m}
                 \mathbf{n} _m \right) \cos \theta \
            \\
            M' _h (K_{sp} , \omega _0) \, p_0 \, e^{i \gamma_1 d}
            e^{i \gamma _m \lvert z \rvert }
            e^{-i \omega _0 t}
        \bigg]
\end{multline}
where $M' _h (K_{sp} , \omega _0)$ is given in appendix
\ref{dip_field}. The surface plasmon field is analogous to a
damped cylindrical wave $\frac{\displaystyle e^{iK_{sp}
\rho}}{\displaystyle \sqrt{K_{sp} \rho}}$ with the same
polarization vector $\mathbf{\hat{\rho}} - \tfrac{\displaystyle
K_{sp} }{\displaystyle\gamma _m} \mathbf{n} _m$, times a factor
$\cos \theta$, making the field vanishing in the direction
perpendicular the dipole (here the $y$ direction) and more intense
in the dipole's direction (here the $x$ direction).

\section{Discussion}

In this section, we revisit two fundamental issues in the field of
surface plasmons~: field confinment and large density of states.
Confinment and superresolution are related to the existence of
wavevectors with a modulus much larger than $\omega/c$. In this
respect the choice of the proper dispersion relation plays a key
role as one has a cut-off wavevector whereas the other predicts no
limit for the dispersion relation. Is there a limit to the
resolution~? Is there a limit to the local density of states~?

\subsection{Super resolution}

Let us first discuss the issue of resolution when imaging with a
surface plasmon driven at frequency $\omega$ by an external
source. Recent experiments on far-field optical microscopy
\cite{Zayats} launched a debate \cite{Zayats2,Drezet} about the
role of surface plasmons in super-resolution imaging effects. In
Ref.~[\onlinecite{Zayats}] the dispersion curve with the
asymptotic behaviour has been invoked to stress the role of
surface plasmons in the image formation with nano-resolution.
 The resolution was estimated to be
$\lambda_{sp}/2$. Therefore, if the dispersion curve with the
asymptotic behaviour is chosen, there seems to be no diffraction
limit and only the amplitude decay of surface plasmon due to Ohmic
losses in the metal limits the resolution. The effect of the
back-bending of surface plasmon dispersion discussed in
Ref.~[\onlinecite{Drezet}] limits the surface plasmon wavelength
$2\pi/\mathfrak{Re}K_{sp}$ and therefore, the resolution. Clearly,
both dispersion relations do not lead to the same conclusion and a
prescription to choose one or the other is needed. Let us consider
a situation where a surface plasmon is excited locally by a
stationary monochromatic field. From section 3, we know that it is
valid to use a representation with fixed amplitudes using modes
with complex wavevectors and real frequencies. This implies that
the dispersion relation with real frequency (with backbending) is
relevant. It follows that there is a cut-off spatial frequency.
Indeed, as $K_x$ may be complex, the propagation term
$\exp(iK_xx)$ introduces damping. In the case of a lossy medium,
damping may be due to losses. However, even for a non-lossy medium
($K_{sp}$ is real), $K_x=(K_{sp}^2-K_y^2)^{1/2}$ can be imaginary.
This occurs when $K_y$ exceeds the value $K_{sp}$. This situation
is the 2D analog of the evanescent waves with wavevector $K$
larger than $\omega/c$ that cannot propagate in a vacuum. Clearly,
$K_{sp}$ is a cut-off frequency and the propagation term
$\exp(iK_xx)$ works as a low-pass filter that prevents propagation
of fields associated with spatial frequencies larger than
$K_{sp}$. When dealing with lossy media, it is the real part of
$K_{sp}$ that specifies the cut-off spatial frequency. It is seen
in Fig.\ref{dispcurve} that  it is limited by the backbending of
the dispersion relation.

In summary, when discussing imaging using stationary monochromatic
surface plasmons, the relevant representation is based on modes
with a complex wavector and a real frequency given by
Eq.(\ref{FourOptSP}). This corresponds to a dispersion relation
that has a backbending. It follows that the resolution is limited
by the cut-off spatial frequency given by the maximum value of
$\mathfrak{Re} K_{sp}$.

\subsection{Local density of states}

Let us now discuss the Local Density of States (LDOS). The density
of states (DOS) is a quantity that plays a fundamental role in
many domains. In particular, it allows to derive all thermodynamic
properties of a system. In the case of an interface, the surface
modes are confined close to the interface so that it is useful to
introduce the Local Density of States (LDOS) that depends on the
distance to the interface\cite{Joulain,ProgSurfSci}. It allows to
account for the huge increase of energy density close to an
interface when surface waves are excited
\cite{Shchegrov,ProgSurfSci}. It also plays a key role in defining
the lifetime of a single emitter close to an
interface\cite{Chance, Henkel, KociakPRL,KociakN,Dereux}. In this
context, the increase of the projected LDOS is usually normalized
by the LDOS in a homogeneous medium (e.g. a vacuum) yielding the
so-called Purcell factor. It is well-known in solid state physics
that the density of states can be derived from the dispersion
relation. More specifically, the DOS increases at a frequency
$\omega$ when the dispersion relation is flat at that particular
frequency. A quick look at Fig.\ref{dispcurve} shows that
different dispersion relations seem to predict different LDOS.
While Fig.\ref{dispcurve}(b) predicts a very large peak at
$\omega_{sp}/\sqrt{2}$ due to the asymptote and no states above
this frequency, Fig.\ref{dispcurve}(a) predicts a smaller peak and
a non zero LDOS between $\omega_{sp}/\sqrt{2}$ and $\omega_{sp}$.
Again, we see that a prescription is needed to choose the right
dispersion relation.

A standard procedure to derive the DOS in the reciprocal space is
based on the periodic boundary conditions. Assuming a surface of
side $L$, the wavevector takes the form
$\mathbf{K}=n_x\frac{\displaystyle 2\pi}{\displaystyle L}
\mathbf{\hat{x}}+n_y\frac{\displaystyle 2\pi}{\displaystyle L}
\mathbf{\hat{y}}$. In the plane $K_x,K_y$, a mode has an area
$4\pi^2/L^2$. It follows that the number of modes per unit area in
$d^2\mathbf{K}$ is given by $d^2\mathbf{K}/4\pi^2$. When
performing this analysis, both $K_x$ and $K_y$ are real. Thus the
relevant representation uses real wavevectors and complex
frequencies. The corresponding dispersion relation has no
backbending and therefore presents a singularity. This is in
agreement with another approach of the LDOS based on the use of
the Green's tensor that predicts an asymptotic behaviour
proportional to $1/(z^3
\vert\epsilon+1\vert^2)$\cite{Joulain,ProgSurfSci}. Of course,
this divergence is non physical. It is related to the modelling of
the medium using a continuous description of the metal. This model
cannot be valid on an atomic scale. Before reaching the atomic
scale, non-local effects must be taken into account.

\section{Conclusion}

The purpose of this work is to clarify several issues regarding
surface plasmons on flat surfaces. The first issue deals with the
mode representation of the surface plasmon field. We have shown
that a surface plasmon field can be represented as a sum of modes
with either a complex wavevector or a complex frequency. We have
shown that a representation using complex frequencies is well
adapted to fields excited by pulses and that a representation
using complex wavevectors is well adapted to stationary
monochromatic fields excited in a finite area. The latter
representation provides a rigorous formula that can be used to
analyse the diffraction of a stationary surface plasmon field.
This should be very useful in order to develop a surface plasmon
Fourier optics framework. This formula clearly shows that the
maximum value of $\mathfrak{Re} K_{sp}$ is a cut-off spatial
frequency that gives an upper limit to the resolution or
confinment  that can be obtained using surface plasmons. As a by
product, we have derived the form of the surface plasmon excited
by a dipole located below the interface. Finally, we have
discussed how to choose the dispersion relation (with or without
backbending) depending on the issue. To illustrate this procedure,
we have shown that there is a resolution limit given by the
maximum value of the wavevector at the backbending point. We have
also shown that the local density of states should be analysed
using the dispersion relation with a real wavevector. This yields
a LDOS that diverges close to the interface in agreement with the
result obtained from the Green's tensor approach.

\appendix
\section{Calculations of the surface plasmon field}
\label{field_calc}
For the plane interface system, it is convenient to use the
representation due to Sipe \cite{Sipe} that consists of a
decomposition over elementary plane waves. We consider the
interface such as the lower medium $z<0$ is denoted as medium 1
and the upper medium $z>0$ is medium $2$. We use the dyadic
notation for the tensor. For instance, the $s$-component of the
electric field is given by
$\hat{\mathbf{s}}\hat{\mathbf{s}}\mathbf{E}=\hat{\mathbf{s}}(\hat{\mathbf{s}}\cdot\mathbf{E})$.
Using the notations of Eq. \eqref{FourG}, we have for the field in
the lower half-space $z<0$ and currents below $z$ ($z'<z$)~:
\begin{multline}
\tensor{g} (\mathbf{K}, z, z', \omega) =
    \tensor{g} _0 (\mathbf{K}, z, z', \omega)
\\
    +\frac{i}{2\gamma_1}
\left[
\hat{\mathbf{s}}r_s\hat{\mathbf{s}}+\hat{\mathbf{p}}_1^-r_p\hat{\mathbf{p}}_1^+
\right] e^{-i\gamma_1 z'} e^{-i\gamma_1 z} \label{tensor}
\end{multline}
where $\tensor{g} _0 (\mathbf{K}, z, z', \omega)$ denotes the
Fourier transform of the Green's tensor of an infinite space
filled by medium $1$.

For the field in the upper half-space ($z>0$) and currents still in the lower half-space ($z'<0$),
one has~:
\begin{multline}
\tensor{g}(\mathbf{K},z, z', \omega) =\frac{i}{2\gamma_1} \left[
\hat{\mathbf{s}}t_s\hat{\mathbf{s}}+\hat{\mathbf{p}}_2
t_p\hat{\mathbf{p}}_1^+ \right] e^{-i\gamma_1 z'} e^{i\gamma_2 z}
\label{tensorabove}
\end{multline}

The Fresnel reflection and transmission factors are given by~:
\begin{align}
r_s &= \frac{\gamma_1-\gamma_2}{\gamma_1+\gamma_2} & r_p &=
\frac{\gamma_1\epsilon_2-\gamma_2\epsilon_1}{\gamma_1\epsilon_2+\gamma_2\epsilon_1}
\label{rp}\\ t_s &= \frac{2 \gamma _1}{\gamma_1+\gamma_2}& t_p &=
\frac{2 \gamma _1 \sqrt{\epsilon _1}\sqrt{ \epsilon _2}
}{\gamma_1\epsilon_2+\gamma_2\epsilon_1} \label{tp}
\end{align}
$\gamma _m = \sqrt{\epsilon _m
\left(\frac{\displaystyle\omega}{\displaystyle c}\right)^2 - K^2}
$ for $m=1,2$ is chosen so that $\mathfrak{Im} \gamma _m > 0$.
This way, using $\epsilon _m (-\omega ^\ast) = \epsilon ^\ast _m
(\omega)$, one has $\gamma _m (K^\ast, -\omega ^\ast) = -\gamma
^\ast _m (K, \omega)$. The square roots of dielectric constants
$\sqrt{\epsilon _m}$ are chosen so that $\mathfrak{Re}
\sqrt{\epsilon _m} \geqslant 0$.
$\hat{\mathbf{s}}=\mathbf{K}\times \mathbf{z}/K$,
$\hat{\mathbf{p}}_1^{\pm}=(K\mathbf{z}\mp\gamma_1\mathbf{K}/K)/(k_0\sqrt{\epsilon_1})$
and $\hat{\mathbf{p}} _2 = (K\mathbf{z} - \gamma_2
\mathbf{K}/K)/(k_0\sqrt{\epsilon_2})$, $k_0=\omega/c$ and $K =
\sqrt{K_x ^2 + K_y ^2}$ is chosen so that $\mathfrak{Im} K >0$ or
$\mathfrak{Im} K = 0$ and $\mathfrak{Re} K > 0$. This way, one has
$\hat{\mathbf{p}} (-\mathbf{K}, -\omega ^\ast) = -
\hat{\mathbf{p}} (\mathbf{K}, \omega) ^\ast$ when $\mathbf{K}$ is
real and $\omega$ complex and $\hat{\mathbf{p}} (-\mathbf{K}
^\ast, -\omega) = \hat{\mathbf{p}} (\mathbf{K}, \omega) ^\ast$
when $\mathbf{K}$ is complex and $\omega$ real.

We extend the definition of $\tensor{g} (\mathbf{K}, z, z', \omega)$ to complex values of $\omega$
and assume that the denominator of the Fresnel coefficients $r_p$ and $t_p$ (whose
nullity is equivalent to Eq. \eqref{reldisp})
has two roots
$\omega _{sp}$ and $-\omega _{sp} ^\ast$~:
\begin{equation}
\frac{1}{\gamma_1 \epsilon_2 + \gamma_2 \epsilon_1}
    = \frac{C(K, \omega)}{(\omega - \omega _{sp})(\omega + \omega _{sp} ^\ast )}
\end{equation}
    with
$\mathfrak{Im} \omega _{sp} <0$.
$\tensor{g}$ then features two poles at $\omega _{sp}$ and $-\omega _{sp} ^\ast$.
The residues of $\tensor{g}$ at these poles can be calculated
with $\tensor{f}_{\widetilde{\omega}} (\mathbf{K}, z, z') =
    \lim _{\omega \rightarrow \widetilde{\omega}} \left[ (\omega - \widetilde{\omega})\tensor{g} (\mathbf{K}, z, z', \omega) \right]$
where $\widetilde{\omega}$ denotes $\omega _{sp}$ or $-\omega _{sp} ^\ast $. It comes~:
\begin{multline} \label{f_wsp}
    \tensor{f}_{\omega _{sp}} (\mathbf{K}, z, z')
        = i\frac{ \gamma_1 \epsilon_2}{ k_0 \sqrt{\epsilon _1}}
            \frac{C(K, \omega _{sp})}{2 \mathfrak{Re} \omega _{sp}}
             (\mathbf{\hat{K}} - \frac{K}{\gamma _m} \mathbf{n}_m)
            \hat{\mathbf{p}} _1 ^+\\
            e^{-i \gamma_1 z'} e^{i \gamma_m \lvert z \rvert }
\end{multline}
where
$\mathbf{n}_m$ denotes $-\hat{\mathbf{z}}$ for $z<0$ and $\hat{\mathbf{z}}$ for $z>0$,
$\gamma _m$ denotes $\gamma_1$ for $z<0$ and $\gamma _2$ for $z>0$.
$\gamma_1$ and $\gamma_2$ depends on $K$ and $\omega _{sp}$ and
$\hat{\mathbf{p}} _1 ^+$ on $\mathbf{K}$ and $\omega _{sp}$.

Using $\epsilon _m (-\omega ^\ast) = \epsilon _m ^\ast (\omega)$,
$\gamma _m (K, -\omega ^\ast) = -\gamma _m ^\ast (K, \omega)$, and
$\hat{\mathbf{p}}_1 ^+ (\mathbf{K}, -\omega ^\ast) = - \hat{\mathbf{p}}_1 ^+ (-\mathbf{K}, \omega) ^\ast$,
we have~:
\begin{equation}  \label{f_mwsps}
    \tensor{f}_{-\omega _{sp} ^\ast} (\mathbf{K}, z, z') =
    -\tensor{f}_{\omega _{sp}} ^\ast (-\mathbf{K}, z, z')
\end{equation}

Using Eq. \eqref{fieldsp}, \eqref{G_w} and \eqref{f_wsp}, one can find the amplitudes in Eq. \eqref{fieldw}~:
\begin{multline}
    E (\mathbf{K}, t) =  -\mu _0\frac{\gamma_1 \epsilon_2}{k_0\sqrt{\epsilon _1}}
    \frac{C(K, \omega _{sp})}{2 \mathfrak{Re} \omega _{sp}}
        \int \mathrm{d} ^2 \mathbf{r}'
        e^{-i \mathbf{K} \cdot \mathbf{r}'} \\
        \int _{-\infty} ^{0} \mathrm{d} z'
        e^{-i \gamma_1 z'}
        \int _{-\infty} ^{t} \mathrm{d} t'
        e^{i \omega _{sp} t'}
        \hat{\mathbf{p}} _1 ^+ . \frac{\partial \mathbf{j}}{\partial t'} (\mathbf{r}', t')
\end{multline}

We now extend the definition of $\tensor{g} (\mathbf{K}, z, z',
\omega)$ to complex values of $K_x$ and assume\footnote{This is
the case for a vacuum/metal (Drude model) interface.} that the
denominator of the Fresnel coefficients $r_p$ and $t_p$ has two
roots $K_{\text{x, sp}} = \sqrt{K _{sp} ^2  - K_\text{y} ^2}$ and
$-K_{\text{x, sp}}$~:
\begin{equation}
\frac{1}{\gamma _1 \epsilon _2 + \gamma_2 \epsilon _1} =
    \frac{\gamma _1 \epsilon _2 - \gamma_2 \epsilon _1}{\epsilon _1 ^2 - \epsilon _2 ^2}
    \frac{1}{(K_\text{x} - K_\text{x, sp})(K_\text{x} + K_\text{x, sp})}
\end{equation}
    with
$\mathfrak{Im} K_{\text{x, sp}} > 0$,
and where $K_{sp}$ depends on $\omega$ and is given by the dispersion relation.
$\tensor{g}$ then features two poles at $K_{\text{x, sp}}$ and $-K_{\text{x, sp}}$.

The residues of $\tensor{g}$ at these poles can be calculated
with $\tensor{f}_{\widetilde{K_\text{x}}} (K_{\text{y}}, z, z', \omega) =
    \lim _{K_\text{x} \rightarrow \widetilde{K_\text{x}}} \left[ (K_\text{x} - \widetilde{K_\text{x}})\tensor{g} (\mathbf{K}, z, z', \omega) \right]$
where $\widetilde{K_\text{x}}$ denotes $K_\text{x, sp}$ or $-K_\text{x, sp} $. It comes~:

\begin{multline}
\label{f_kxsp}
    \tensor{f}_{K_\text{x,sp}} (K_{\text{y}}, z, z', \omega)
    =\frac{i}{2 K_{\text{x, sp}}}\frac{ \gamma_1 \epsilon _2}{k_0\sqrt{\epsilon _1}}
    \frac{\gamma _1 \epsilon _2 - \gamma_2 \epsilon _1}{\epsilon _1 ^2 - \epsilon _2 ^2}
         \\
        (\hat{\mathbf{K}} ^+ - \tfrac{K}{\gamma _\text{m}} \mathbf{n} _\text{m})
        \hat{\mathbf{p}} _1 ^+ (K_\text{x,sp} \mathbf{\hat{x}} + K_\text{y} \mathbf{\hat{y}},\omega)
        e^{-i\gamma_1 z'}
        e^{i\gamma_\text{m} \lvert z \rvert}
\end{multline}
\begin{multline}
\label{f_mkxsp}
    \tensor{f}_{-K_\text{x,sp}} (K_{\text{y}}, z, z', \omega)
    =-\frac{i}{2 K_{\text{x, sp}}} \frac{ \gamma_1 \epsilon _2}{k_0\sqrt{\epsilon _1}}
    \frac{\gamma _1 \epsilon _2 - \gamma_2 \epsilon _1}{\epsilon _1 ^2 - \epsilon _2 ^2}
        \\
        (\hat{\mathbf{K}} ^- - \tfrac{K}{\gamma _\text{m}} \mathbf{n} _\text{m})
        \hat{\mathbf{p}} _1 ^+ (-K_\text{x,sp} \mathbf{\hat{x}} + K_\text{y} \mathbf{\hat{y}},\omega)
        e^{-i\gamma_1 z'}
        e^{i\gamma_\text{m} \lvert z \rvert}
\end{multline}
where $\hat{\mathbf{K}} ^+ = \frac{\displaystyle K_{\text{x, sp}}
\mathbf{\hat{x}} + K_\text{y} \mathbf{\hat{y}}}{\displaystyle
K_{sp}}$, $\hat{\mathbf{K}} ^- = \frac{\displaystyle -K_{\text{x,
sp}} \mathbf{\hat{x}} + K_\text{y} \mathbf{\hat{y}}}{\displaystyle
K_{sp}}$, and the other notations are defined above.

Using Eq. \eqref{fieldsp}, \eqref{G_Kx}, \eqref{G_mKx}, \eqref{f_kxsp} and \eqref{f_mkxsp}, one can find the amplitudes
in Eq. \eqref{fieldKx}~:
\begin{multline}
E _{>} (K _{\text{y}} , \omega, x)
    =\mu _0\frac{1}{2 K_{\text{x, sp}}}\frac{\gamma_1 \epsilon _2}{k_0\sqrt{\epsilon _1}}
    \frac{\gamma _1 \epsilon _2 - \gamma_2 \epsilon _1}{\epsilon _1 ^2 - \epsilon _2 ^2}
         \\
        \int _{-\infty} ^{x} \mathrm{d}x' e^{-i K_\text{x,sp} x'}
        \int \mathrm{d}y' e^{-iK_\text{y} y'}
        \int _{-\infty} ^{0} \mathrm{d}z'
        e^{-i\gamma_1 z'}
        \int \mathrm{d}t' e^{i \omega t'} \\
        \hat{\mathbf{p}} _1 ^+ (K_\text{x,sp} \mathbf{\hat{x}} + K_\text{y} \mathbf{\hat{y}},\omega)
        .  \frac{\partial \mathbf{j}}{\partial t'} (\mathbf{r}', t')
\label{Edr}
\end{multline}
\begin{multline}
E _{<} (K _{\text{y}} , \omega, x)
    =\mu _0\frac{1}{2 K_{\text{x, sp}}}\frac{\gamma_1 \epsilon _2}{k_0\sqrt{\epsilon _1}}
    \frac{\gamma _1 \epsilon _2 - \gamma_2 \epsilon _1}{\epsilon _1 ^2 - \epsilon _2 ^2}
         \\
        \int _{x} ^{\infty} \mathrm{d}x' e^{-i K_\text{x,sp} x'}
        \int \mathrm{d}y' e^{-iK_\text{y} y'}
        \int _{-\infty} ^{0} \mathrm{d}z'
        e^{-i\gamma_1 z'}
        \int \mathrm{d}t' e^{i \omega t'} \\
        \hat{\mathbf{p}} _1 ^+ (-K_\text{x,sp} \mathbf{\hat{x}} + K_\text{y} \mathbf{\hat{y}},\omega)
        .  \frac{\partial \mathbf{j}}{\partial t'} (\mathbf{r}', t')
\end{multline}

\section{Surface plasmon field of a dipole}
\label{dip_field}
The currents associated with the dipole
$\mathbf{p}_0 e^{-i\omega_0 t}
$
at a distance $d$ above the interface
are given by
$\mathbf{j} (\mathbf{r}, t) = 2 \mathfrak{Re} \left[ e^{-i\omega_0 t} (-i) \omega _0 \mathbf{p}_0\right] \delta(\mathbf{r} - (-d) \mathbf{\hat{z}})$.
Using the form of the surface plasmon field given by Eq. \eqref{FourOptSP},
one can compute the amplitude with this expression and Eq. \eqref{Edr}~:
\begin{multline}
    E _> (K_y, \omega) =
        - \frac{\gamma _1 \epsilon _2 - \gamma _2 \epsilon _1}{\epsilon _1 ^2 - \epsilon _2 ^2}
        \frac{\gamma _1 \epsilon _2}{k_0\sqrt{\epsilon _1}} \frac{1}{2 K_{\text{x,sp}}}
        e^{i \gamma _1 d} \mu _0 \omega _0 ^2
        \\
        \mathbf{\hat{p}} _1 ^+ .
            \left[
                2\pi \delta (\omega - \omega _0) \mathbf{p}_0
                + 2\pi \delta (\omega + \omega _0) \mathbf{p}_0 ^\ast
            \right]
\end{multline}

Hence, using Eq. \eqref{FourOptSP} and the properties
$K(-\mathbf{K} ^\ast ) = -K ^\ast (\mathbf{K})$,
$K_{\text{x,sp}} (K_y, -\omega) = -K_{\text{x,sp}} ^\ast (K_y, \omega)$,
$\gamma _m (K ^\ast, -\omega) = -\gamma _m ^\ast (K, \omega)$,
$\epsilon _m (-\omega) = \epsilon _m ^\ast (\omega)$
and the definition of $\mathbf{\hat{p}} _1 ^+$, it comes~:
\begin{multline}
    \mathbf{E_{\rm m}} =-
         \mathfrak{Re} \Bigg[
            e^{i \gamma _m \lvert z \rvert} e^{-i \omega _0 t}
            \frac{\gamma _1 \epsilon _2 - \gamma _2 \epsilon _1}{\epsilon _1 ^2 - \epsilon _2 ^2}
            \frac{\gamma _1 \gamma _2}{\epsilon _0}
            \\
            \int \frac{\mathrm{d} K_y}{2\pi}
                \frac{e^{i \mathbf{K}\cdot\mathbf{r}}}{K_{\text{x,sp}}}
                (\mathbf{\hat{K}} - \tfrac{\displaystyle K}{\displaystyle \gamma _m} \mathbf{\hat{n}} _m)
                (\mathbf{\hat{K}} - \tfrac{\displaystyle K}{\displaystyle \gamma _1} \mathbf{\hat{z}})
                . \mathbf{p} _0
        \Bigg]
\end{multline}

By derivating $ \int \frac{\displaystyle \mathrm{d}
K_y}{\displaystyle 2\pi} \frac{\displaystyle e^{i
\mathbf{K}\cdot\mathbf{r}}}{\displaystyle K_{\text{x,sp}}}
    = \tfrac{1}{2} H_0^{(1)} (K_{sp} \rho) $
with respect to $x$ or $y$ one finds $ \int
\frac{\displaystyle\mathrm{d} K_y}{\displaystyle2\pi}
\frac{\displaystyle e^{i \mathbf{K}\cdot\mathbf{r}}}{\displaystyle
K_{\text{x,sp}}} \mathbf{\hat{K}}
    = \tfrac{i}{2} H_1^{(1)} (K_{sp} \rho) \mathbf{\hat{\rho}} $
and $ \int \frac{\displaystyle\mathrm{d} K_y}{\displaystyle 2\pi}
\frac{\displaystyle e^{i \mathbf{K}\cdot\mathbf{r}}}{\displaystyle
K_{\text{x,sp}}}
        \mathbf{\hat{K}} \mathbf{\hat{K}}
    = \tfrac{1}{2} \left[
                [H_1^{(1)} (K_{sp} \rho)]' \mathbf{\hat{\rho}} \mathbf{\hat{\rho}}
                + \frac{\displaystyle H_1^{(1)} (K_{sp} \rho)}{\displaystyle K_{sp} \rho} \mathbf{\hat{\theta}} \mathbf{\hat{\theta}}
            \right] $.
Using these relations and the value of $\mathbf{p} _0$, with $M(K,
\omega) = - \frac{\displaystyle\gamma _1 \gamma _2}{\displaystyle4}
\frac{\displaystyle\gamma _1 \epsilon _2 - \gamma _2 \epsilon
_1}{\displaystyle\epsilon _1 ^2 - \epsilon _2 ^2} $, we get
Eq.\eqref{vertdip} and \eqref{horizdip}. According to
Eq.\eqref{FourOptSP}, the field found in Eq.\eqref{vertdip} and
\eqref{horizdip} only apply in the $x>0$ half-space. Using
symmetries arguments, one find easily that they also apply in the
$x<0$ half-space.

Eq.\eqref{vertdip} and \eqref{horizdip} can then be simplified
using the asymptotical form of the Hankel functions $H_n(z)
\rightarrow \sqrt{\frac{\displaystyle 2}{\displaystyle\pi z}}
e^{iz - \tfrac{1}{2}\pi i (n + \tfrac{1}{2})}$. Denoting $M' _v
(K_{sp} , \omega _0) = - \sqrt{\frac{\displaystyle
2}{\displaystyle\pi}} e^{-i\frac{\pi}{4}} M(K_{sp} , \omega _0)
\tfrac{\displaystyle K_{sp} }{\displaystyle\gamma _1\epsilon _0}
$, we obtain Eq.\eqref{vertdips}. Using also the property of the
Hankel functions $H' _1 (z) = H_0 (z) - \tfrac{\displaystyle
1}{\displaystyle z} H_1(z)$ and denoting $M' _h (K_{sp} , \omega
_0) = \sqrt{\frac{\displaystyle 2}{\displaystyle\pi}}
e^{-i\frac{\pi}{4}} M(K_{sp} , \omega _0) \frac{\displaystyle
1}{\displaystyle\epsilon _0} $, we obtain Eq.\eqref{horizdips}.

\begin{acknowledgments}
 This work was supported by the French National Research Agency (ANR)
 through Carnot Leti funding
 and by the French Ministry of Defense through
 a grant from the \emph{Direction g\'en\'erale de l'armement} (DGA).

\end{acknowledgments}

\end{document}